# A semianalytical approach for determining the nonclassical mechanical properties of materials

Mohammad Reza Zamani[1,2,3] [*] and Usef Jafaraghaei[4]

[1]*Center for High Technology Materials, University of New Mexico, Albuquerque, NM 87106, USA.*
[2]*Mechanical Engineering Department, University of New Mexico, Albuquerque, NM 87131, USA.*
[3]*Electerical and Computer Engineering Department, University of New Mexico, Albuquerque, NM 87131, USA.*
[4]*Mechanical Engineering Department, Sirjan University of Technology, Kerman, Iran.*

**Abstract**

In this article, a semianalytical approach for demonstrating elastic waves' propagation in nanostructures has been presented based on the modified couple-stress theory including acceleration gradients. Using the experimental results and atomic simulations, the static and dynamic length scales were calculated for several materials, zinc oxide (ZnO), silicon (Si), silicon carbide (SiC), indium antimonide (InSb), and diamond. To evaluate the predicted static and dynamic length scales as well as the presented model, the natural frequencies of a beam in addition to the phase velocity and group velocity of Si were studied and compared with the available static length scales, estimated using strain-gradient theory without considering acceleration gradients. These three criteria, natural frequency, phase velocity, and group velocity, show that the presented model is dynamically stable even for larger wavevector values. Furthermore, it is explained why the previous works, which all are based on the strain-gradient theory without acceleration gradients, predicted very small values for the static length scale in the longitudinal direction rather than the static length scale in the transverse directions.

## I. INTRODUCTION

Phonons, which describe vibrational motions in condensed matter, are collective excitations in the periodic elastic arrangement of atoms [1]. Phonon propagation in materials, either single crystals or polycrystalline materials, has been used widely for characterizing materials because it illustrates the vibrational modes as well as the atomic interactions in materials. Studying phonon dispersions has become more and more important and favored for characterizing materials as scientific instruments and measurement techniques, such as dispersive-phonon imaging [2], picosecond laser ultrasonic [3], infrared spectroscopy [4], neutron spectroscopy [5], Raman spectroscopy [6], and Brillouin spectroscopy [7], were developed leading to a significant improvement in accurately measuring the phonon modes propagating in solids. These achievements resulted in revealing unbounded information regarding materials' properties, such as mechanical [8-9], thermal and thermoelectric [10-11], electrical and flexo-electrical [12-13], and chemical [14], suitable for engineering the materials' functionalities. Therefore, it is crucial to employ phonon dispersions to understand fully materials' properties to optimize their performance or tune their properties.

One of the most interesting applications of phonon dispersions is to understand the mechanical behaviors of materials on nanoscales. Indeed, it was shown that to demonstrate precisely the phonon dispersions at the center of the first Brillouin zone based on the solid mechanics theories, very complicated solid mechanics formulas including several material constants, also called nonclassical properties, are required [9, 15]. These complicated solid mechanics formulas had not been explained based on the classical continuum mechanics definition until the nonclassical solid mechanics theories, such as couple-stress [16] and strain-gradient theories [17], were developed. Depending on the dimensions of the structures, the nonclassical mechanical properties have significant influences on mechanical behaviors of structures, especially when dimensions are nanometers or submicrometers, or negligible influences on the mechanical behaviors of the structures when the dimensions are millimeters or larger [8]. Equivalently, the classical continuum mechanics, which first proposed for bulk dimensions, fails to illustrate the mechanical behaviors of nanostructures because it ignores the nonclassical mechanical properties (also called length scales) at nanoscales [18-20].

Moreover, it has been shown over the years that classical continuum mechanics fails to demonstrate precisely the effects of surface energy [21], discrete nature of the materials [22], presence of micro- and nanocracks [23], internal strains, and quantum confinements in the nanostructures [24]. These effects can be only investigated and explained using the nonclassical theories [8]. Therefore, it is crucial to implement nonclassical solid mechanics theories including nonclassical mechanical properties to illustrate the mechanical performance of nanostructures. However, to be theoretically and experimentally consistent, the first step is to establish a semianalytical approach, a combination of experimental and theoretical techniques.

---

[*] Corresponding author: mrzamani@unm.edu, Tel.: +1 (618)-305-9636

Phonon dispersions can be derived using well-established quantum mechanics approaches, such as density-functional perturbation theory [25] and density-functional theory [26], or lattice-dynamic approaches, such as deformable-ion model, polarizable bond charges model, and rigid valence shell model [27-30]. To the knowledge of the authors, DiVincenzo [9] considered dispersive elastic constants for the first time to illustrate the deviation between the experimental results, the ballistic phonon imaging experiment, and the classical continuum mechanics as well as the lattice-dynamics simulation. Later, Maranganti and Sharma [8] used the second strain-gradient theory without acceleration gradients in conjugation with lattice-dynamics simulations and molecular-dynamics simulations to connect the dispersive elastic constants to the length scales. In contrast to Maranganti and Sharma [8], Jakata and Every [31] only focused on the experimental data to find the nonclassical mechanical properties with the same approach. It should be mentioned the difference between the results of Maranganti and Sharma [8] and Jakata and Every [31] could be due to lack of a connection between the theoretical and experimental approaches because each approach has some extended approximations and errors. A more complicated analysis was proposed by Wang et. al. [32], where they described the wave dispersion using a gradient-enriched shell theory and gradient-enriched Timoshenko theory. Their approach resulted in an imaginary group velocity besides unusual cut-off frequencies that are inconsistent with other work [15]. However, Askes and Aifantis [33] represented a revision of Wang et. al. [32] model leading to an improvement in the simulation results. Furthermore, in conjugation with the strain-gradient approach, there is always a contradiction regarding the sign of gradient terms along with concerns of uniqueness and stability versus the ability to describe dispersive acoustic waves' propagation. This controversial discussion was professionally reviewed in Askes and Aifantis [15] while it was not evaluated using other nonclassical theories, such as couple-stress theory [16] or modified couple-stress theory (MCST) [34], and being combined with both atomic simulations and experimental data to resolve this contradiction fully.

Consequently, here we predict the nonclassical mechanical constants of several materials based on a semianalytical approach. To do so, we rigorously derive the most general forms of the phonon dispersions in the framework of MCST including the acceleration gradients. Then, the phonon-dispersions curves of the materials were calculated using atomic simulations. To fit the atomic simulation results with the derived phonon dispersions, we use the experimental values for the classical constants and then we find the nonclassical constants through fitting the atomic simulation results. To compare the present results of this approach with the earlier works, which were determined based on strain-gradient theory without considering acceleration gradients, we calculated the phase velocity and group velocity of these materials in addition to the natural frequency of a simply–simply supported beam (SSSB) with different static length scale to radius ratios.

## II. EQUILIBRIUM EQUATIONS

According to the MCST of Yang et. al. [18] the strain energy density is a function of both strain (conjugated with stress) and curvature (conjugated with couple stress). It then follows that the strain energy $U$ of a deformed linear elastic material occupying region $\Omega$ is given by

$$U = \frac{1}{2}\int_\Omega (\sigma_{kl}\varepsilon_{kl} + m_{kl}\chi_{kl})d\Omega, \tag{1}$$

where the stress tensor, $\sigma_{ij}$, strain tensor, $\varepsilon_{ij}$, deviatoric part of the couple-stress tensor, $m_{ij}$, and symmetric curvature tensor, $\chi_{ij}$, are, respectively, defined by

$$\sigma_{kl} = C_{ijkl}\varepsilon_{ij}, \tag{2}$$

$$m_{kl} = \beta_{ijkl}\chi_{ij} \tag{3}$$

$$\boldsymbol{\varepsilon} = \frac{1}{2}\left(\nabla\mathbf{u} + (\nabla\mathbf{u})^T\right), \tag{4}$$

$$\boldsymbol{\theta} = \frac{1}{2}curl(\mathbf{u}), \tag{5}$$

$$\boldsymbol{\chi} = \frac{1}{2}\left(\nabla\boldsymbol{\theta} + (\nabla\boldsymbol{\theta})^T\right). \tag{6}$$

The $C_{ijkl}$ and $\beta_{ijkl}$ are the material constants expressing the classical constants and nonclassical constants, respectively. Furthermore, **u** is the displacement vector; and **θ** is the rotation vector. To consider the effects of the anisotropic properties of the materials, the materials constants are considered in the general form. However, to be consistent with the original form of the MCST and for simplicity, only one static length scale is considered, such that $\beta_{ijkl} = 8l_s^2 C_{ijkl}$, where $l_s$ is the static length scale. The total kinetic energy is given by

$$T = \frac{1}{2}\int_\Omega \left(\rho \dot{u}_i \dot{u}_i + \rho l_d^2 \dot{u}_{i,j} \dot{u}_{i,j}\right) d\Omega. \tag{7}$$

This form of the kinetic energy, which is a simplified version of the Mindlin [17] definition, has attracted special attention especially in the dynamical analysis of nanostructures. That is why this definition of the kinetic energy has an extra length scale, named the dynamical length scale, so it eliminates the dynamical instabilities [15]. It should be mentioned, because the focus is on studying the phonon-dispersion relations, the external loadings are ignored. Therefore, the total energy stored in the materials is

$$\Pi = U - T. \tag{8}$$

To derive the governing equations, the minimum energy principle is applied. Therefore,

$$\partial \Pi = \partial(U - T) = 0 \quad \Rightarrow \quad \partial U = \partial T. \tag{9}$$

Applying the variational theorem to Eq. (1) and using the symmetry property of the $C_{ijkl}$ and $\beta_{ijkl}$, one obtains

$$\partial U = \int_\Omega \left(C_{ijkl}\varepsilon_{kl}\partial\varepsilon_{ij} + \beta_{ijkl}\chi_{kl}\partial\chi_{ij}\right) d\Omega. \tag{10}$$

Using the definitions of the strain tensor, symmetric curvature tensor, and microrotations vector provided in Eq. (4) to Eq. (6), their variational forms are, respectively, determined by

$$\partial \varepsilon_{ij} = \frac{1}{2}\left(\partial u_{i,j} + \partial u_{j,i}\right), \tag{11}$$

$$\partial \chi_{ij} = \frac{1}{2}\left((\partial \theta_i)_{,j} + (\partial \theta_j)_{,i}\right), \tag{12}$$

$$\partial \theta_i = \frac{1}{2}e_{imn}(\partial u_n)_{,m}. \tag{13}$$

Substituting Eqs. (11)–(13) into Eq. (10), it can be rearranged as

$$\partial U = \int_t \int_\Omega \left(C_{ijkl}\varepsilon_{kl}(\partial u_i)_{,j} + \left[\left(\beta_{ijkl}\chi_{kl}\partial\theta_i\right)_{,j} - \frac{1}{2}e_{mji}\beta_{mnkl}\chi_{kl,n}(\partial u_i)_{,j}\right]\right) d\Omega dt \tag{14}$$

Applying integration-by-parts on the first and third terms in Eq. (14) gives

$$\partial U = \int_t \int_\Omega \left(\begin{array}{l}\left[-C_{ijkl}\varepsilon_{kl,j} + \frac{1}{2}e_{mji}\beta_{mnkl}\chi_{kl,nj}\right]\partial u_i \\ + \left[C_{ijkl}\varepsilon_{kl}\partial u_i - \frac{1}{2}e_{mji}\beta_{mnkl}\chi_{kl,n}\partial u_i + \beta_{ijkl}\chi_{kl}\partial\theta_i\right]_{,j}\end{array}\right) d\Omega dt \tag{15}$$

The last step to simplify Eq. (15) is to apply the divergence theorem to the second bracket. The divergence theorem converts the integral over the volume to an integral over the surface as follows

$$\partial U = \int_t \int_\Omega \left(\left[-C_{ijkl}\varepsilon_{kl,j} + \frac{1}{2}e_{mji}\beta_{mnkl}\chi_{kl,nj}\right]\partial u_i\right) d\Omega dt$$
$$+ \int_t \int_S \left[C_{ijkl}\varepsilon_{kl}\partial u_i - \frac{1}{2}e_{mji}\beta_{mnkl}\chi_{kl,n}\partial u_i + \beta_{ijkl}\chi_{kl}\partial\theta_i\right]n_j dS dt \tag{16}$$

To find the variational form of the kinetic energy, the same process can be repeated for Eq. (7). Therefore,

$$\partial T = \int_t \int_\Omega \left(\rho \dot{u}_i \frac{\partial}{\partial t}(\partial u_i) + \rho l_d^2 \dot{u}_{i,j}\frac{\partial}{\partial t}(\partial u_i)_{,j}\right) d\Omega dt \tag{17}$$

Applying the integration-by-parts technique with respect to time, Eq. (17) can be rearranged as

$$\partial T = \int_t \int_\Omega \left( -\rho \ddot{u}_i \partial u_i - \rho l_d^2 \ddot{u}_{i,j} (\partial u_i)_{,j} \right) d\Omega dt + \int_\Omega \left( \rho \dot{u}_i \partial u_i + \rho l_d^2 \dot{u}_{i,j} (\partial u_i)_{,j} \right) d\Omega \qquad (18)$$

Again, the integration-by-parts technique must be applied to the second term of the first integral of Eq. (18) with respect to the spatial derivative. Thus,

$$\partial T = \int_t \int_\Omega \left[ \left( -\rho \ddot{u}_i + \rho l_d^2 \ddot{u}_{i,jj} \right) \partial u_i + \left( \rho l_d^2 \ddot{u}_{i,j} \partial u_i \right)_{,j} \right] d\Omega dt + \int_\Omega \left( \rho \dot{u}_i \partial u_i + \rho l_d^2 \dot{u}_{i,j} (\partial u_i)_{,j} \right) d\Omega \qquad (19)$$

The last step to derive the variational form of the kinetic energy is to apply the divergence theorem to the last terms under the integrals of Eq. (19). Eventually, one gets

$$\partial T = \int_t \int_\Omega \left( -\rho \ddot{u}_i + \rho l_d^2 \ddot{u}_{i,jj} \right) \partial u_i d\Omega dt + \int_t \int_S \left( \rho l_d^2 \ddot{u}_{i,j} \partial u_i \right) n_j dS dt + \int_\Omega \rho \dot{u}_i \partial u_i d\Omega + \int_S \rho l_d^2 \dot{u}_{i,j} \partial u_i n_j dS \qquad (20)$$

Inserting Eq. (16) and Eq. (20) into Eq. (9), the equations of motion will be achieved as follows,

$$C_{ijkl}\varepsilon_{kl,j} - \frac{1}{2} e_{mji} \beta_{mnkl} \chi_{kl,nj} = \rho \ddot{u}_i - \rho l_d^2 \ddot{u}_{i,jj} \qquad \text{or } \partial u_i = 0 \text{ on } \Omega \qquad (21a)$$

$$C_{ijkl}\varepsilon_{kl} n_j - \frac{1}{2} e_{mji} \beta_{mnkl} \chi_{kl,n} n_j = \rho l_d^2 \ddot{u}_{i,j} n_j \qquad \text{or } \partial u_i = 0 \text{ on } S \qquad (21b)$$

$$\beta_{ijkl} \chi_{kl} n_j = 0 \qquad \text{or } \partial \theta_i = 0 \text{ on } S \qquad (21c)$$

Eqs. (21) are the most general form of the equations of motion in the scope of the MCST including the acceleration gradients. It should be emphasized that, because here the focus is to derive the mechanical constants using an atomic simulation, the equilibrium equations are free from external traction terms. However, the external tractions can be readily added into Eqs. (21) based on the domain that they are exerting on. Furthermore, for an isotropic linear elastic material, the material constants will be replaced with

$$C_{ijkl} = \lambda \delta_{ij} \delta_{kl} + \mu \left( \delta_{ik} \delta_{jl} + \delta_{il} \delta_{kj} \right) \qquad (22a)$$

$$\beta_{ijkl} = \left( \lambda \delta_{ij} \delta_{kl} + \mu \left( \delta_{ik} \delta_{jl} + \delta_{il} \delta_{kj} \right) \right) l_s^2 \qquad (22b)$$

Substituting Eqs. (22) into Eqs. (21), the following equations of motion for an isotropic linear elastic material can be obtained

$$\lambda u_{j,ji} + \mu \left( u_{j,ji} + u_{i,jj} \right) + l_s^2 \mu \left( u_{j,jikk} - u_{i,jjkk} \right) = \rho \ddot{u}_i - \rho l_d^2 \ddot{u}_{i,jj} \qquad \text{or } \partial u_i = 0 \text{ on } \Omega \qquad (23a)$$

$$\lambda u_{k,k} \delta_{ij} n_j + \mu \left( u_{i,j} + u_{j,i} \right) n_j - \mu l_s^2 e_{mji} e_{mpq} u_{q,pkk} n_j = \rho l_d^2 \ddot{u}_{i,j} n_j \qquad \text{or } \partial u_i = 0 \text{ on } S \qquad (23b)$$

$$\chi_{ij} n_j = 0 \qquad \text{or } \partial \theta_i = 0 \text{ on } S \qquad (23c)$$

Ignoring the dynamic length scale in Eqs. (23), the MCST's governing equations of motions can be achieved as proposed by several authors [17, 34-35].

### III. DISPERSION RELATIONS

In this section, the phonon-dispersion relations are derived for both longitudinal and transverse phonon modes. To do so, a longitudinal plane wave (LPW) ($u_1 \neq 0$, $u_2 = u_3 = 0$) and transverse plane waves (TPWs) ($u_1 = u_3 = 0$, $u_2 \neq 0$ and $u_1 = u_2 = 0$, $u_3 \neq 0$) are considered and implemented into Eq. (21a). Consequently, the dispersion relations for the longitudinal and transverse modes can be determined as

LPW: $\begin{aligned} u_1 &= A_1 \exp[i(kx - \omega t)] \\ u_2 &= u_3 = 0 \end{aligned}$ $\qquad \dfrac{\omega_L^2}{V_{L_e}^2} = \dfrac{k^2}{1 + l_d^2 k^2} \qquad V_{L_e} = \sqrt{\dfrac{C_{1111}}{\rho}},$ (24a)

TPW: $\begin{aligned} u_2 &= A_2 \exp[i(kx - \omega t)] \\ u_1 &= u_3 = 0 \end{aligned}$ $\qquad \dfrac{\omega_{T_2}^2}{V_{T_{2e}}^2} = \dfrac{1 + Rl_s^2 k^2}{1 + l_d^2 k^2} k^2 \qquad V_{T_{2e}} = \sqrt{\dfrac{C_{2121}}{2\rho}} \text{ and } R = \dfrac{4C_{3131}}{C_{2121}},$ (24b)

TPW: $\begin{aligned} u_3 &= A_3 \exp[i(kx - \omega t)] \\ u_1 &= u_2 = 0 \end{aligned}$ $\qquad \dfrac{\omega_{T_3}^2}{V_{T_{3e}}^2} = \dfrac{1 + Rl_s^2 k^2}{1 + l_d^2 k^2} k^2 \qquad V_{T_{3e}} = \sqrt{\dfrac{C_{3131}}{2\rho}} \text{ and } R = \dfrac{4C_{2121}}{C_{3131}}.$ (24c)

It should be mentioned that Eqs. (24) carry much interesting information. First, the static length scale does not contribute to the longitudinal dispersion relation, Eq. (24a). This fact can be explained from both solid mechanics and atomic points of view. From the solid mechanics view, that is why the curvature tensor is deviatoric, its diagonal terms associated with the longitudinal terms are zero. From the atomic view, the longitudinal phonons have higher energy than the transverse phonons. However, this fact does not mean the longitudinal terms are immune to dispersion because the dynamic length scale appears in the longitudinal phonon-dispersion relation, which highlights the critical role of the dynamic length scale in demonstrating the phonon-dispersion relation. This conclusion is compatible with the prediction of the all previous works using the strain gradient theory that they calculated the static length scale for the longitudinal modes to be much smaller than the static length scale for the transverse modes. Second, Eqs. (24b) and (24c) explain the effects of the unit cell asymmetry on the transverse dispersion relations with fewer materials constants comparing the strain-gradient theory. For example, for materials with the orthorhombic crystal structure, such as agrinierite and krennerite, because all the lattice constants are different, the transverse phonon-dispersion relations in different directions are different. Lastly, all dispersion relations, Eqs. (24), are dynamically stable and they suggest the phonon frequencies to be real and finite regardless of the $l_i k$ ($i = s, d$) values.

Eqs. (25) represent the acoustic waves' propagation in $k$-(110) and $k$-(111) directions, respectively. In Eqs. (25), for simplicity, the $j$-index is summation from 1 to 3. It should be mentioned that the remaining acoustic modes can be readily determined using Eqs. (25).

$$
\begin{aligned}
&u_1 = u_1(x, y, t) \\
&u_2 = u_3 = 0
\end{aligned}
\quad
\omega^2 = \frac{\varsigma + \zeta}{\rho\left(1 + l_d^2\left(k_1^2 + k_2^2\right)\right)}
\quad
\begin{aligned}
\varsigma &= \frac{1}{2}\left[\begin{array}{l} 2C_{1111}k_1^2 + (2C_{1112} + C_{1121})k_1 k_2 \\ + (C_{1212} + C_{1221})k_2^2 \end{array}\right], \\
\zeta &= \left[C_{1313}k_1^2 k_2^2 + 2C_{3132}k_1 k_2^3 + C_{3232}k_2^4\right]l_s^2.
\end{aligned}
\tag{25a}
$$

$$
\begin{aligned}
&u_1 = u_1(x, y, z, t) \\
&u_2 = u_3 = 0
\end{aligned}
\quad
\omega^2 = \frac{\xi + \gamma}{\rho\left(1 + l_d^2\left(k_1^2 + k_2^2 + k_3^2\right)\right)}
\quad
\begin{aligned}
\xi &= \frac{1}{2}\left[\begin{array}{l} 2C_{1j11}k_1 + (C_{1j12} + C_{1j21})k_2 \\ + (C_{1j13} + C_{1j31})k_3 \end{array}\right]k_j, \\
\gamma &= \left[\begin{array}{l} \left(\begin{array}{l} C_{2j12}k_1 k_3 + C_{2j13}k_1 k_2 \\ + C_{2j21}k_1 k_3 + 2C_{2j22}k_1 k_3 \\ + C_{2j23}(k_3^2 - k_2^2) - C_{2j31}k_1 k_2 \\ + C_{2j32}(k_3^2 - k_2^2) - 2C_{2j33}k_2 k_3 \end{array}\right)k_j k_3 \\ -\left(\begin{array}{l} C_{3j12}k_1 k_3 + C_{3j13}k_1 k_2 \\ + C_{3j21}k_1 k_3 + 2C_{3j22}k_1 k_3 \\ + C_{3j23}(k_3^2 - k_2^2) - C_{3j31}k_1 k_2 \\ + C_{3j32}(k_3^2 - k_2^2) - 2C_{3j33}k_2 k_3 \end{array}\right)k_j k_2 \end{array}\right]l_s^2.
\end{aligned}
\tag{25b}
$$

## IV. RESULTS AND DISCUSSIONS
### A. Mechanical properties of ZnO, Si, SiC, InSb, and diamond

In this section, mechanical properties of several materials, ZnO, Si, SiC, InSb, and diamond, are studied using a semianalytical approach. First, the phonon-dispersion relations of the materials were determined using an atomic simulation in the Virtual NanoLab Atomistic ToolKit (VNL-ATK) software [37]. Then, the results of the simulations were analyzed by the definitions of the phonon dispersions derived in Eqs. (24). Each phonon-dispersion relation consists of at least two material constants, one classical and one nonclassical. According to Eqs. (24), the longitudinal mode includes a classical constant and the dynamic length scale while the transverse mode includes both static and dynamic length scales in addition to the classical constants. Therefore, using the available experimental results for the classical constants, the longitudinal phonon-dispersion mode first was fitted with the atomic simulation results to find the dynamic length scale. Then, using the classical constants of the experiments in conjunction with this dynamic length scale, the transverse phonon-dispersion modes were fitted with the atomic simulation results to achieve the static length scale as well. Because the procedure is the same for all materials, here we explain the procedure only for silicon (Si). Fig. 1 shows the phonon dispersion of Si.

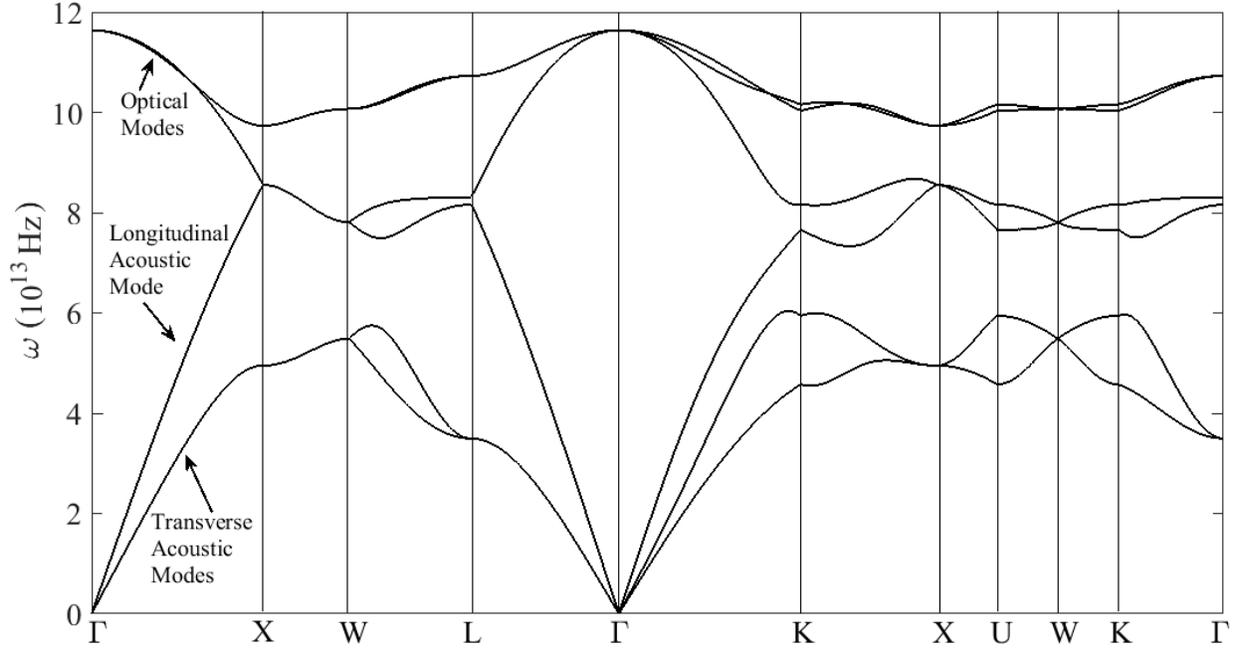

FIG. 1. Phonon dispersion of Si calculated using the Tersoff potential provided by Erhart and Karsten [38].

Table I shows the estimated values for the length scales and a comparison with the literature. The experimental data used for the classical constants of Si are provided from Feng et. al. [39]. It is noteworthy that, in addition to the completely different approaches used in the present work and the earlier works, here the phonon-dispersion relations were fitted for $k$-wavevectors up to 0.1 while they were fitted for much smaller values of $k$-wavevectors, for example, $|k| < 0.05$ in Maranganti and Sharma [8].

TABLE I. Estimated length scales for Si.

|  | Transverse | | | | Longitudinal | | | |
|---|---|---|---|---|---|---|---|---|
|  | This work | | | | This work | | | |
| Static length scale (nm) | 0.21 | 0.225[a] | 0.07[b] | 0.12[c] | 0 | 0.13[a] | 0.234[b] | 0.12[c] |
| Dynamic length scale (nm) | 0.11 | - | - | - | 0.11 | - | - | - |

[a] Reference [8].
[b] Reference [31].
[c] Reference [3].

Table II provides the predicted values for the dynamic and static length scales of Si, SiC, ZnO, InSb, and diamond. The experimental data used for the classical constants of SiC, ZnO, InSb, and diamond are provided from Kamitani et. al. [40], Feng et. al. [39], Devaux et. al. [41], and Field [42], respectively.

TABLE II. The static and dynamic length scales of Si, SiC, ZnO, InSb, and diamond.

| | Materials | | | | | | |
|---|---|---|---|---|---|---|---|
| | Si | SiC | ZnO | InSb | InSb* | Diamond | Diamond* |

| | | | | | 0.24[1b], | | 0.025[1a], |
|---|---|---|---|---|---|---|---|
| Static length scale (nm) | 0.21 | 0.30 | 0.14 | 0.34 | 0.37[2b] | 0.60 | 0.192[2a] |
| Dynamic length scale (nm) | 0.11 | 0.29 | 0.34 | 0.21 | - | 0.38 | - |

* Superscripts 1 and 2 represent the values for longitudinal and transverse directions, respectively.
a Reference [8].
b Reference [31].

**B. Phase velocity and group velocity**

In this section, the phase and group velocities of Si were studied as a criterion for comparing the results of this work and the previous results. That is why the phase velocity and group velocity are strong tools for studying the dispersion modes inside a nanostructure. The phase of an acoustic wave in a nanostructure is the rate at which the phase of the wave propagates. Therefore, the phase velocity is defined as

$$v_{p_i} = \frac{\omega_i}{k}, \quad i = L, T, \tag{26}$$

where $\omega_i$ is the angular frequency, $L$ and $T$ stand for longitudinal and transverse modes, respectively. Implementing Eqs. (24) and the estimated length scales in Table I into Eq. (26), the normalized longitudinal and transverse phase velocities of the Si were calculated and shown in Fig. 2. Furthermore, the normalized longitudinal and transverse phase velocities of the Si were calculated using the strain-gradient theory.

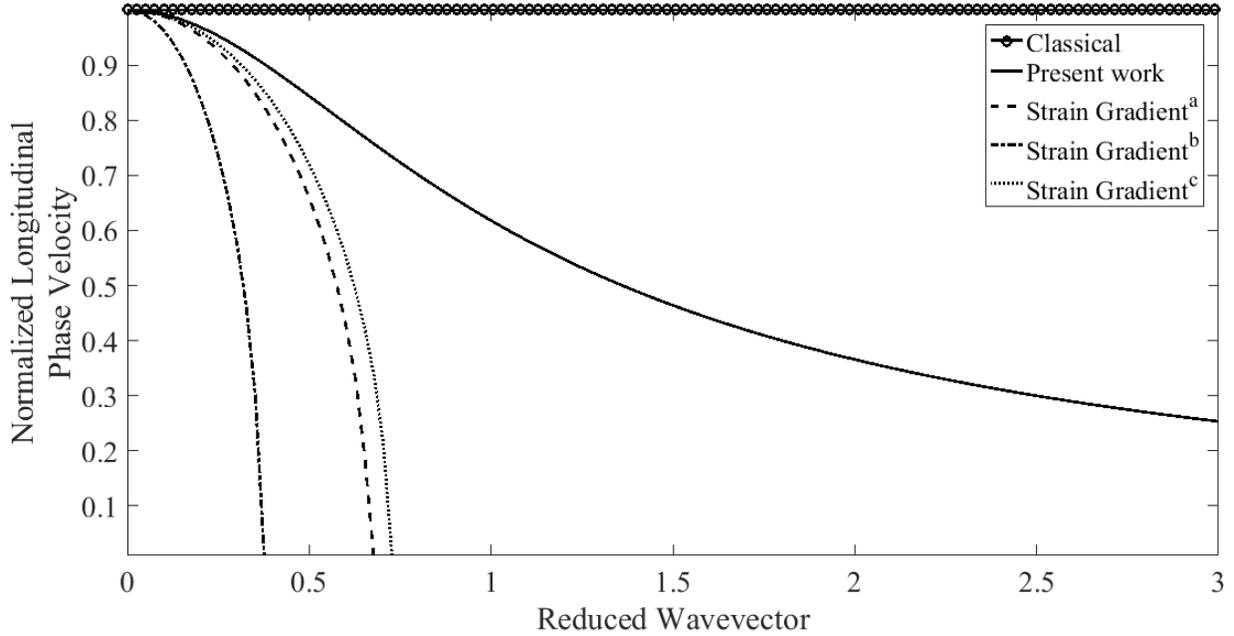

(a): The normalized longitudinal phase velocity.

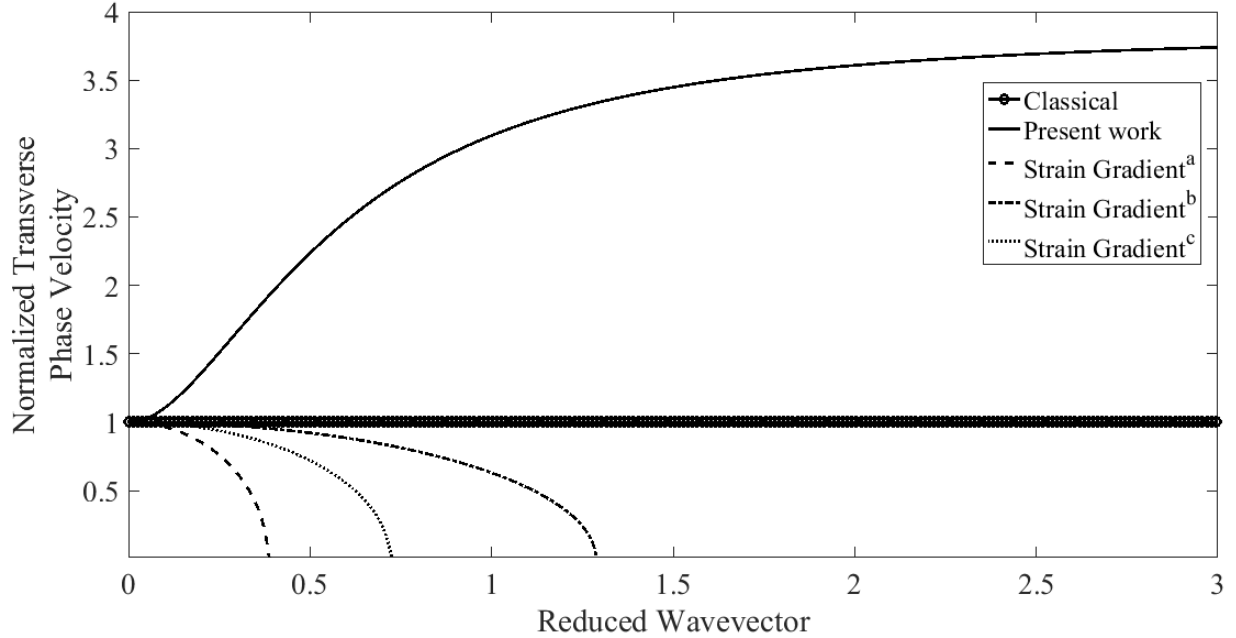

(b): The normalized transverse phase velocity.

FIG. 2. The normalized longitudinal phase velocity, (a), and the normalized transverse phase velocity, (b), of Si. a, b, and c curves were plotted according to the strain-gradient theory and predicted mechanical constants in Maranganti and Sharma [8], Jakata and Every [31], and Hao and Maris [3], respectively.

As can be seen from Fig. 2, the previous works based on the strain-gradient models, which do not consider acceleration gradients, show unstable behaviors [15]. That is why these models predict the square of the longitudinal and transverse phase velocities to grow unboundedly toward negative infinity as the wavevector increases, resulting in imaginary phase velocities. This unrealistic behavior is due to the sign paradox of the strain gradient terms as well as neglecting the effects of acceleration gradients and considering very small wavevectors for predicting the mechanical constants of Si. Considering very small wavevectors for fitting the atomic simulations or experimental data caused those works to correctly predict the longitudinal and transverse phase velocities for extremely small wavevectors, but failed for larger wavevectors [15].

In addition to the longitudinal and transverse phase velocities, we calculated the longitudinal and transverse group velocities of Si by the same process explained for the longitudinal and transverse phase velocities. According to the definition of the group velocity, which is the velocity with which the overall shape of the wave amplitude propagates, the longitudinal and transverse group velocities can be calculated as

$$v_{g_i} = \frac{\partial \omega_i}{\partial k}, \quad i = L, T. \tag{27}$$

Fig. 3 shows the normalized longitudinal and transverse group velocities inside the Si. The explanation for the group velocities' behavior is the same as for the phase velocities.

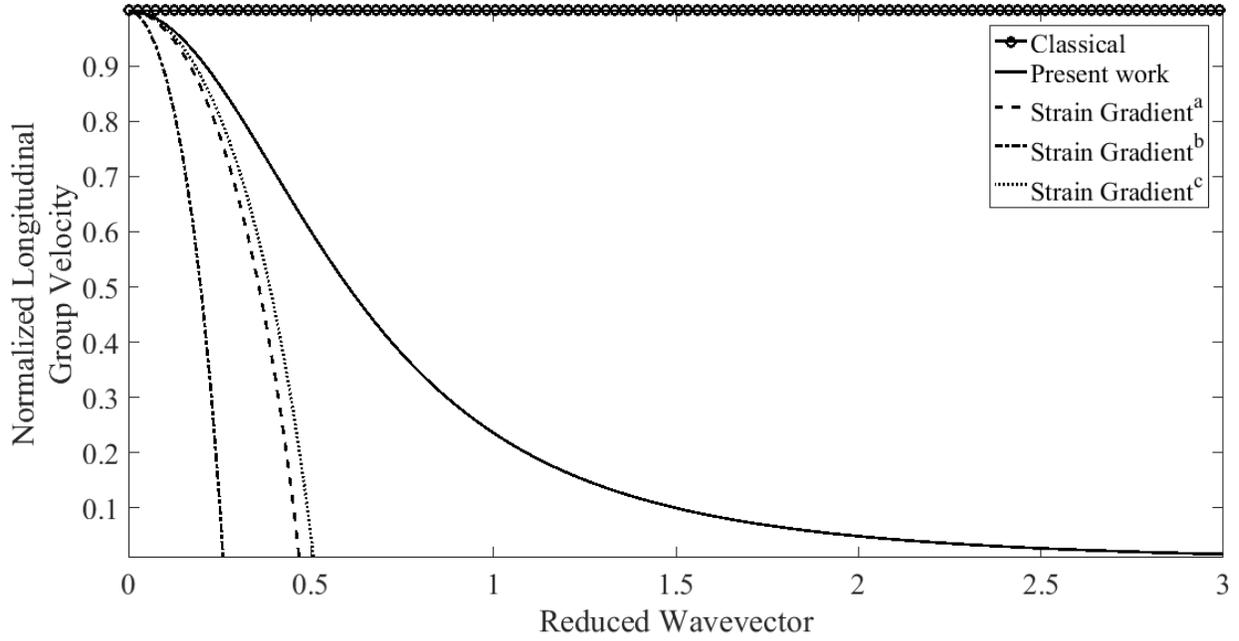

(a): The normalized longitudinal group velocity.

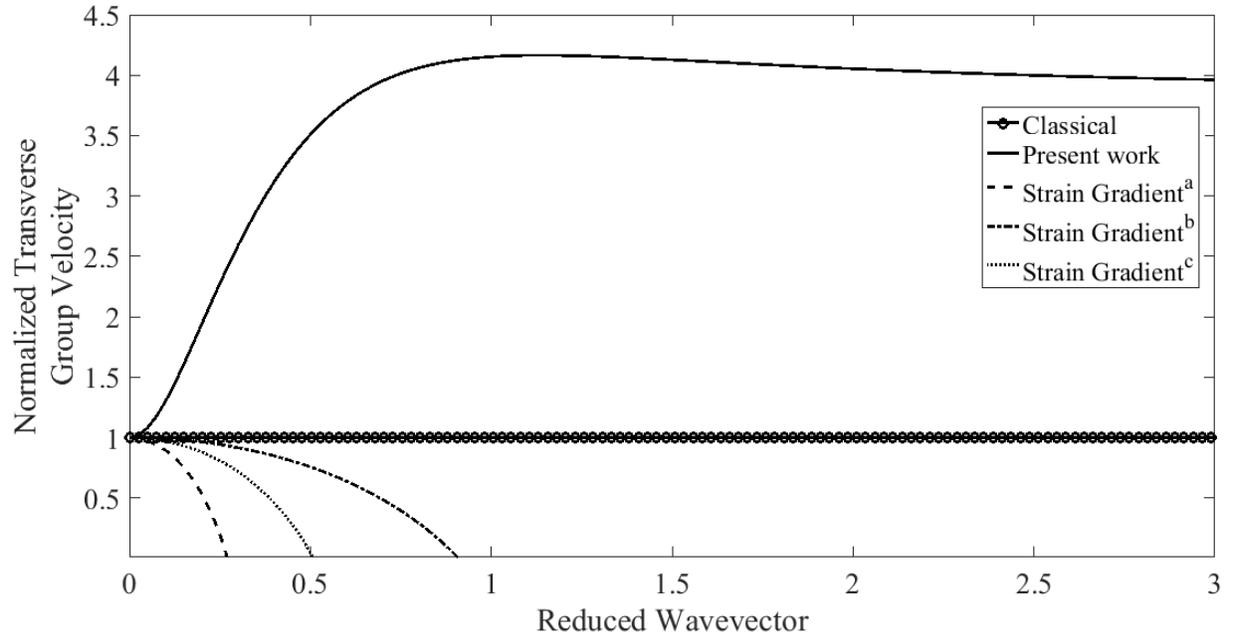

(b): The normalized transverse group velocity.

FIG. 3. The normalized longitudinal phase velocity, (a), and the normalized transverse phase velocity, (b), of Si. a, b, and c curves were plotted according to the strain-gradient theory and predicted mechanical constants in Ref. [8], Ref. [31], and Ref. [3], respectively.

## C. Natural frequency

Lastly, in this section to evaluate the MCST with acceleration gradient and strain-gradient theory for demonstrating the dynamical behavior of nanostructures, we calculate the natural frequencies of an SSSB. The general forms of the governing equation of motions and the boundary conditions as well as initial conditions of a beam can be determined using Hamilton's principle as mentioned in the first section, Eq. (9). For a one-dimensional beam, the displacement components can be considered, Dym and Shames [43],

$$u_x = -z\frac{dw(x)}{dx}, \quad u_y = 0, \quad u_z = w(x), \tag{28}$$

where $u_x$, $u_y$, and $u_z$ are the displacement components in the $x$, $y$, and $z$ directions, respectively. Here, it is assumed that the beam is along the $x$-axis in the $x$–$z$ plane. Substituting the displacement components into Eq. (9), for free vibration of a beam, one gets

$$\begin{aligned}\partial(U-T) &= \partial\int_t\int_0^L\left[\frac{1}{2}\left(C_{1111}I + 4C_{1212}l_s^2 A\right)\left(\frac{d^2 w}{dx^2}\right)^2 - \frac{1}{2}\left(\rho A\left(\frac{dw}{dt}\right)^2 + \rho l_d^2 A\left(\frac{d}{dx}\left(\frac{dw}{dt}\right)\right)^2\right)\right]dxdt \\ &= \int_t\int_0^L\left[\left(C_{1111}I + 4C_{1212}l_s^2 A\right)\frac{d^4 w}{dx^4} + \rho A\frac{d^2 w}{dt^2} - \rho l_d^2 A\frac{d^2}{dx^2}\left(\frac{d^2 w}{dt^2}\right)\right]\partial w\, dxdt \\ &\quad + \int_t\left[\left(C_{1111}I + 4C_{1212}l_s^2 A\right)\frac{d^3 w}{dx^3}\partial w\right]_0^L dt - \int_0^L\left[\left(\rho A\frac{dw}{dt} - \rho l_d^2 A\frac{d^2}{dx^2}\left(\frac{dw}{dt}\right)\right)\partial w\right]_t dx \\ &\quad + \int_t\left[\left(C_{1111}I + 4C_{1212}l_s^2 A\right)\frac{d^2 w}{dx^2}\partial\left(\frac{dw}{dx}\right)\right]_0^L dt - \int_t\left[\rho l_d^2 A\frac{d}{dx}\left(\frac{dw}{dt}\right)\partial\left(\frac{dw}{dt}\right)\right]_0^L dt = 0.\end{aligned} \tag{29}$$

Consequently, the governing equation of motion, initial conditions, and boundary conditions are given by

Governing equation of motion:

$$\left(C_{1111}I + 4C_{1212}l_s^2 A\right)\frac{d^4 w}{dx^4} + \rho A\frac{d^2 w}{dt^2} - \rho l_d^2 A\frac{d^2}{dx^2}\left(\frac{d^2 w}{dt^2}\right) = 0. \tag{30a}$$

Initial conditions:

$$\left(\rho A\frac{dw}{dt} - \rho l_d^2 A\frac{d^2}{dx^2}\left(\frac{dw}{dt}\right)\right)\partial w\bigg|_t = 0,$$

$$\frac{d}{dx}\left(\frac{dw}{dt}\right)\partial\left(\frac{dw}{dt}\right)\bigg|_t = 0. \tag{30b}$$

Boundary conditions:

$$\begin{aligned}w &= 0 \quad \text{or} \quad \frac{d^3 w}{dx^3} = 0 \\ \frac{dw}{dx} &= 0 \quad \text{or} \quad \frac{d^2 w}{dx^2} = 0\end{aligned} \tag{30c}$$

Eqs. (30) provide the most general forms of the governing equation of motion, initial conditions, and boundary conditions of a beam under the free vibration condition. However, because here our goal is to compare the MCST including acceleration gradients with the strain-gradient theory, we only consider an SSSB, and then we calculate its first natural frequency. For free vibration of an SSSB, it can be shown that the solution is in the form of $\omega(x, t) = X(x)\exp(i\omega t)$. Therefore, the governing equation of motion will be

$$\left(C_{1111}I + 4C_{1212}l_s^2 A\right)\frac{d^4 X}{dx^4} + \rho l_d^2 A\omega^2\frac{d^2 X}{dx^2} - \rho A\omega^2 X = 0. \tag{31}$$

Because Eq. (31) is a fourth-order differential equation with constant coefficients, the general form of its solution is $X(x) \propto \exp(\lambda t)$. Therefore, its characteristic equation is given by

$$\begin{aligned}B\lambda^4 + D\lambda^2 - G &= 0. \\ B = C_{1111}I + 4C_{1212}l_s^2 A, \quad D &= \rho l_d^2 A\omega^2, \quad G = \rho A\omega^2\end{aligned} \tag{32}$$

The solutions of the characteristic equation are

$$\lambda_1, \lambda_2 = \pm j\sqrt{\frac{D+\sqrt{D^2+4BG}}{2B}}$$
$$\lambda_3, \lambda_4 = \pm\sqrt{\frac{-D+\sqrt{D^2+4BG}}{2B}}$$
(33)

Therefore, the $X(x)$ is

$$X(x) = a_1 \sin(\sqrt{\frac{D+\sqrt{D^2+4BG}}{2B}}) + a_2 \cos(\sqrt{\frac{D+\sqrt{D^2+4BG}}{2B}})$$
$$+ a_3 \sinh(\sqrt{\frac{-D+\sqrt{D^2+4BG}}{2B}}) + a_4 \cosh(\sqrt{\frac{-D+\sqrt{D^2+4BG}}{2B}}).$$
(34)

For an SSSB, the four boundary conditions are
$$w(0) = w(L) = 0,$$
$$\frac{d^2w}{dx^2}(0) = \frac{d^2w}{dx^2}(L) = 0.$$
(35)

Implementing Eq. (34) into Eqs. (35) results in a system of four equations that can be written as follows

$$[H(\omega)]_{4\times 4} \begin{Bmatrix} a_1 \\ a_2 \\ a_3 \\ a_4 \end{Bmatrix} = 0.$$
(36)

To have a nonzero solution for $a_i$ ($i =1, 2, 3, 4$) of Eq. (36), the following conditions must be satisfied
$$\det([H(\omega)]) = 0.$$
(37)

Solving Eq. (37) provides all the natural frequencies of the SSSB. The natural frequencies of an SSSB based on the strain gradient can be determined using the same procedure. Here, we avoided calculating the natural frequencies of an SSSB based on strain gradient, the interested readers are referred to Kong et. al. [44] and Papargyri-Beskou et. al. [45] for more details.

Fig. 4 shows the results of normalized values for the first natural frequency of an SSSB in terms of radius to the static length scale ratio. The normalized natural frequencies based on all methods approach each other as the radius to the static length scale ratio increases. For smaller ratios, the present model predicts lower values for the natural frequency because of the existence of the acceleration gradients that act as an extra term in the inertia matrix.

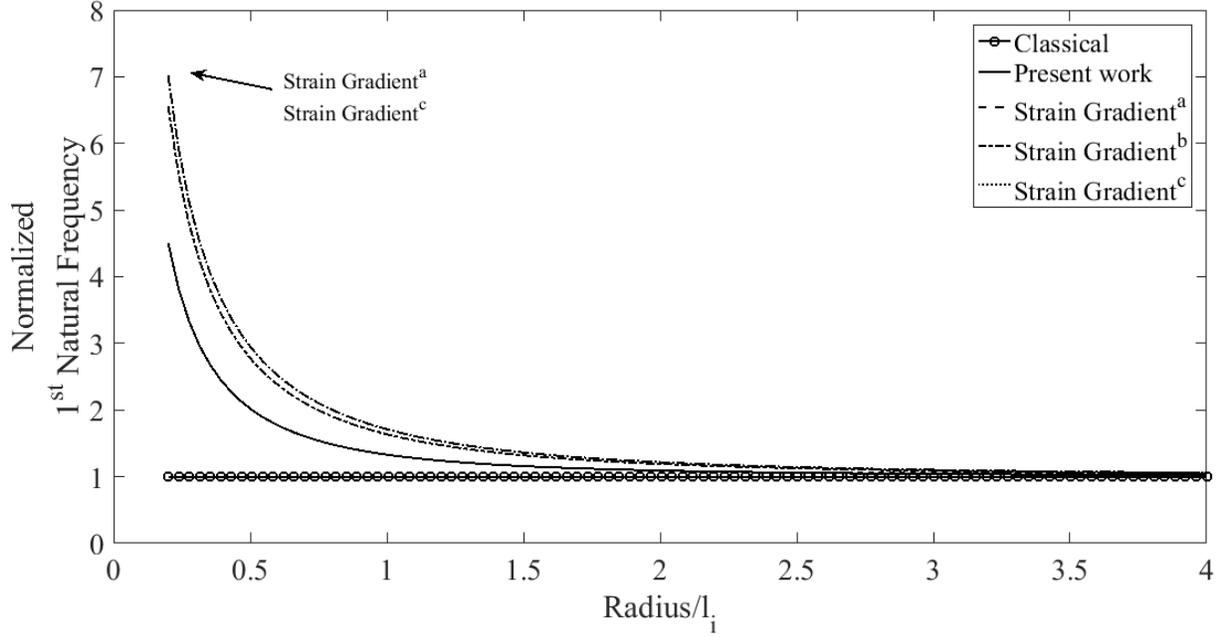

FIG. 4. Normalized first natural frequency of an SSSB based on the strain-gradient theory and the MCST with acceleration gradients, $l_i$ is the static length scale in each theory. The curves are normalized with respect to the first natural frequency according to the classical continuum mechanics. a, b, and c are from Ref. [8], Ref. [31], Ref. [3], respectively.

## V. CONCLUSION

Here, we represented the equation of motion and dispersion relations based on the MCST with acceleration gradients. Using MCST with acceleration gradients for determining the dispersion relations suggests that the static length scale does not contribute to the longitudinal phonon dispersions; however, the dynamic length scale contributes to the phonon-dispersion relations, while both the static and dynamic length scales affect the transverse phonon dispersions. This result is compatible with the strain-gradient theory, where it predicts the static length scale in the longitudinal direction to be very small compared with the static length scale in transverse directions. Combining these formulas with experimental results and atomic simulations, a semianalytical approach, the nonclassical length scales, static and dynamic length scales, were predicted for ZnO, Si, SiC, InSb, and diamond. To evaluate the accuracy of these formulas and technique, the phase velocity, group velocity, and natural frequency of Si SSSB were examined. This approach suggests the phase velocity and group velocity to be finite and real, which proves this approach provides dynamically stable behavior for nanostructures even for large values of wavevector. Furthermore, because of the appearance of the dynamical length scale, this approach predicts the natural frequencies to be somewhat lower than predicted natural frequencies with the strain-gradient approach.